\documentclass[11pt,onecolumn,floatfix,altaffilletter,superscriptaddress,tightenlines,showpacs,showkeys,preprintnumbers,nofootinbib]{revtex4-1}
\pdfoutput=1
\usepackage[colorlinks=true,citecolor=blue,linkcolor=blue,breaklinks=true]{hyperref}
\usepackage{amsmath,amssymb}
\usepackage{mathtools}
\usepackage{epsfig} 
\usepackage{graphicx}        
\usepackage{url}
\usepackage{color}
\usepackage{multirow}
\usepackage{placeins}
\usepackage[dvipsnames]{xcolor}
\usepackage{braket}
\usepackage{float}
\usepackage{slashed}
\usepackage{tikz}
\hypersetup{
  colorlinks=true,
  linkcolor=red, 
  filecolor=magenta,   
  urlcolor=blue,
  citecolor=blue,
}

\allowdisplaybreaks

\bibpunct{[}{]}{,}{n}{}{,}

\def\beq{\begin{equation}}
\def\eeq{\end{equation}}
\def\bea{\begin{eqnarray}}
\def\eea{\end{eqnarray}}
\makeatletter

\begin{document}
\title{PBH-infused seesaw origin of matter and unique gravitational waves}



\author{Debasish Borah}
\email{dborah@iitg.ac.in}
\affiliation{Department of Physics, Indian Institute of Technology Guwahati, Assam 781039, India}
\author{Suruj Jyoti Das }
\email{suruj@iitg.ac.in}
\affiliation{Department of Physics, Indian Institute of Technology Guwahati, Assam 781039, India}
\author{Rome Samanta}
\email{samanta@fzu.cz}
\affiliation{CEICO, Institute of Physics of the Czech Academy of Sciences, Na Slovance 1999/2, 182 21 Prague 8, Czech Republic}
\author{Federico R. Urban}
\email{federico.urban@fzu.cz}
\affiliation{CEICO, Institute of Physics of the Czech Academy of Sciences, Na Slovance 1999/2, 182 21 Prague 8, Czech Republic}


\begin{abstract}
The Standard Model, extended with three right-handed (RH) neutrinos, is the simplest model that can explain light neutrino masses, the baryon asymmetry of the Universe, and dark matter (DM). Models in which RH neutrinos are light are generally easier to test in experiments. In this work, we show that, even if the RH neutrinos are super-heavy ($M_{i=1,2,3}>10^9$ GeV)---close to the Grand Unification scale---the model can be tested thanks to its distinct features on the stochastic Gravitational Wave (GW) background. We consider an early Universe filled with ultralight primordial black holes (PBH) that produce a super-heavy RH neutrino DM via Hawking radiation. The other pair of RH neutrinos generates the baryon asymmetry via thermal leptogenesis, much before the PBHs evaporate. GW interferometers can test this novel spectrum of masses thanks to the GWs induced by the PBH density fluctuations. In a more refined version, wherein a $U(1)$ gauge symmetry breaking dynamically generates the seesaw scale, the PBHs also cause observable spectral distortions on the GWs from the $U(1)$-breaking cosmic strings. Thence, a low-frequency GW feature related to DM genesis and detectable with a pulsar-timing array must correspond to a mid- or high-frequency GW signature related to baryogenesis at interferometer scales.


\end{abstract} 
\maketitle

\section{Introduction}
Neutrino oscillation data \cite{Esteban:2020cvm} suggests that the active neutrino masses are tiny, $m_i\sim \mathcal{O}(0.01)$ eV. Type-I seesaw \cite{seesaw1,seesaw2,seesaw3,seesaw4} is the simplest and the most elegant mechanism, wherein, owing to the hierarchy between the Electroweak scale and the scale of Grand Unification (GUT) \cite{gut1,gut2,gut3}, such small masses can be understood. Quantitatively, the Type-I seesaw mechanism indicates that, if the Yukawa couplings are not strongly fine-tuned, then $m_i\simeq \Lambda_{\rm EW}^2/\Lambda_{\rm GUT}$, where $\Lambda_{\rm EW}\simeq 100$ GeV and $\Lambda_{\rm GUT}\simeq 10^{15}$ GeV.  In a renormalizable seesaw Lagrangian, these two scales are introduced with the right-handed (RH) neutrino sterile fields ($N_R$), and via two mass terms: $\mathcal{L}_{\rm mass}\sim  \Lambda_{\rm EW} \overline{L} N_R+\Lambda_{\rm GUT} \overline{N^c_R} N_R$, where $L$ is the Standard Model (SM) lepton doublet. The heavy fields, $N_R$, besides generating light neutrino masses, also decay CP-asymmetrically to produce the Baryon Asymmetry of the Universe (BAU) via leptogenesis \cite{lep1,lep2,lep3,lep4}. While a two-RH-neutrino extension of the SM suffices to address the generation of light neutrino masses and the BAU, it is natural to complete the sterile fermion family with a third RH neutrino, analogously to the other SM leptons. In this case, if cosmologically stable, one of the RH neutrinos could be a Dark Matter (DM) candidate. Therefore, a three-RH-neutrino extension of the SM model provides an elegant and unified explanation of three ``beyond the SM'' problems. Many efforts have been made towards such unifications, see, e.g., \cite{uni1,uni2,uni3,uni4,uni5,uni6,uni7,uni8,uni9,uni10}. Because so far the energy reachable with colliders is only up to a few TeV, in order to develop a testable unification model, we are mostly led to consider light sterile fermions with masses much below $\Lambda_{\rm GUT}$.\\

In this article, we show that a testable unification model is possible even in keeping with the original implementation of the seesaw, i.e., considering all three RH fermions to be super-heavy, close to $\Lambda_{\rm GUT}$. The catch is that, contrary to the previous models, here we propose to search for the signatures of such unification in Gravitational Waves (GW) experiments -- an idea also entertained in Refs.~\cite{sp1,sp2,sp3,sp4,sp5,sp6,sp6,sp7,sp8,sp9,sp10,sp11} that aims to find GW signatures of the high-scale seesaw within various cosmological context.\\

Suppose exotic objects such as ultralight primordial black holes (PBH) with initial monochromatic mass $M_{\rm BH}$ exist prior to the Big Bang Nucleosynthesis (BBN) ($T\gtrsim$ 5 MeV) \cite{bbn1,bbn2,bbn3}. They could be abundant enough to dominate the Universe's energy budget before evaporating via Hawking radiation. A PBH-dominated Universe is associated with two significant consequences. First, while evaporating, the PBHs produce entropy that dilutes any pre-existing relics. Second, if the PBHs dominate, strong and sharply peaked GWs are induced due to PBH density fluctuations \cite{ingwm1,ingw0,ingw1,ingw2,ingw3,ingw4}. For a fixed value of $M_{\rm BH}$, both the amount of produced entropy and the amplitude of the GWs increase with the duration of the PBH-domination epoch. Furthermore, since the Hawking radiation from a black hole is a gravitational phenomenon, along with the SM particles, the PBHs must produce the DM, independently of its non-gravitational interaction. In the case where the PBHs dominate at relatively low temperatures close to the BBN, they produce super-heavy DM ($M_{DM}\gtrsim 10^{10}$ GeV), constituting the  observed DM abundance \cite{shdm0,shdm1,shdm2,shdm3}.\\

Consider now that the DM produced by the PBHs is one of the RH neutrinos ($N_3\equiv N_{\rm DM}$) in the seesaw. For a  scale of leptogenesis, $T_{\rm lepto}\sim M_1\gtrsim 10^9$ GeV \cite{lepover1}, irrespective of whether the masses of $N_{1,2}$ are hierarchical or quasi-degenerate \cite{lepover2}, the BAU is overproduced for a region in the Yukawa parameter space. The PBHs, which evaporate at much lower temperatures, bring the overproduced asymmetry down to the observed value via entropy dilution. Because a large $T_{\rm lepto}$ requires substantial entropy dilution, hence a longer duration of PBH-domination, the amplitude of the induced GWs increases with $T_{\rm lepto}\sim M_1$. In this scenario the DM mass is related to $M_{\rm BH}$ and it determines the peak frequency of the induced GWs. As the RH neutrino masses approach $\Lambda_{\rm GUT}$, the mechanism naturally predicts strong GWs with amplitudes within reach of LIGO \cite{LIGO1,LIGO2}, ET \cite{et}, and CE \cite{ce}. Therefore, a PBH-dominated Universe provides a unique opportunity to test the origin of ordinary matter and the DM in the seesaw, where the amplitude and the peak frequency of the predicted GWs are determined by super-heavy RH neutrino masses.\\

In the following sections, we present the general framework of the model, discuss the model's technicalities and results, and model extensions, before summarising and concluding.\\

\section{General framework and assumptions}
We consider non-rotating PBHs with monochromatic initial mass $M_{\rm BH}<10^9~g$, produced in a radiation-dominated Universe.  If the PBHs dominate the Universe's energy density, they produce only super-heavy DM with masses, e.g., $M_{\rm DM}\simeq 10^{10} \, {\rm GeV}-10^{15}$ GeV  at temperatures $T_{\rm eva}^{\rm DM}\simeq T_{\rm BBN}-100~{\rm GeV}$ \cite{shdm0,shdm1,shdm2,shdm3}. Such super-heavy DM is produced because, as the PBHs evaporate, the Hawking temperature $T_{\rm BH}=(8\pi G M_{\rm BH})^{-1}$  continues to  increase and finally becomes larger than the $M_{\rm DM}$. If the $\rm PBH\rightarrow DM$ process makes up all  the DM that we observe today \cite{Planck}, the following relation holds:
 \bea
M_{\rm DM}\simeq 4.5\times 10^3\left(\frac{M_{\rm BH}}{M_{\rm Pl}}\right)^{-5/2}M_{\rm Pl}^2~~{\rm GeV^{-1}}, \label{hdm}
\eea
where $M_{\rm Pl}=1.22\times 10^{19}$ GeV is the Planck mass.

\begin{figure}
\includegraphics[scale=.4]{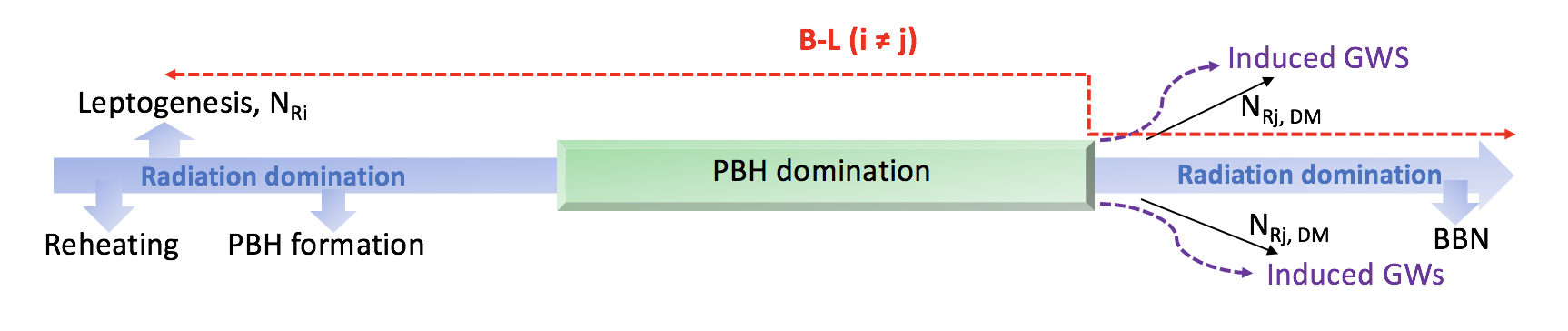}
\caption{A possible timeline for the proposed scenario.}\label{fig1}
\end{figure}

First, we shall consider the simplest scenario where the DM ($N_{i=3, DM}$) is strictly stable and does not interact with any other particle. We then assume that $[T_{\rm Bf},M_{i=1,2}]<T_{\rm RH}$, where $T_{\rm RH}$ and  $T_{\rm Bf}$ are the reheating and PBH formation temperatures respectively. This condition ensures that,  once the Universe reheats after inflation, the PBHs form during radiation domination. Thermal scatterings mediated by Yukawa interactions populate the $N_i$s, which seed baryogenesis via thermal leptogenesis \cite{lep1}. Furthermore, even though the DM does not have any interaction, the PBHs produce $N_{\rm DM}$ via Hawking radiation\footnote{PBHs also produce $N_{1,2}$. Nonetheless, the PBH evaporation temperature corresponding to the correct dark matter relic is lower than the sphaleron freeze-out temperature. Therefore, the lepton asymmetry produced by $N_{1,2}$ will not be processed into a baryon asymmetry.}. We shall neglect any pre-existing DM relic \cite{infdm1,infdm2,infdm3}, but the extension to that case is straightforward. A possible timeline of the proposed mechanism is shown in Fig.\ref{fig1}.

Thus, in this paper, we consider non-thermal DM production from PBH evaporation and matter generation via thermal leptogenesis. In principle, the mechanism could work also if leptogenesis proceeds non-thermally, but thermal scatterings populate the DM density. However, the PBH mass window for non-thermal leptogenesis, $M_{BH}\in \left[0.1~{\rm g}-10~{\rm g}\right]$ \cite{frd2}, would produce very high-frequency induced GWs---see Eq.(\ref{gul2}) below---which would be beyond the reach of current and planned GW interferometers, making this scenario less predictive and testable.

\section{Black hole signatures for the origin of matter in the seesaw}
\begin{figure}
\includegraphics[scale=.5]{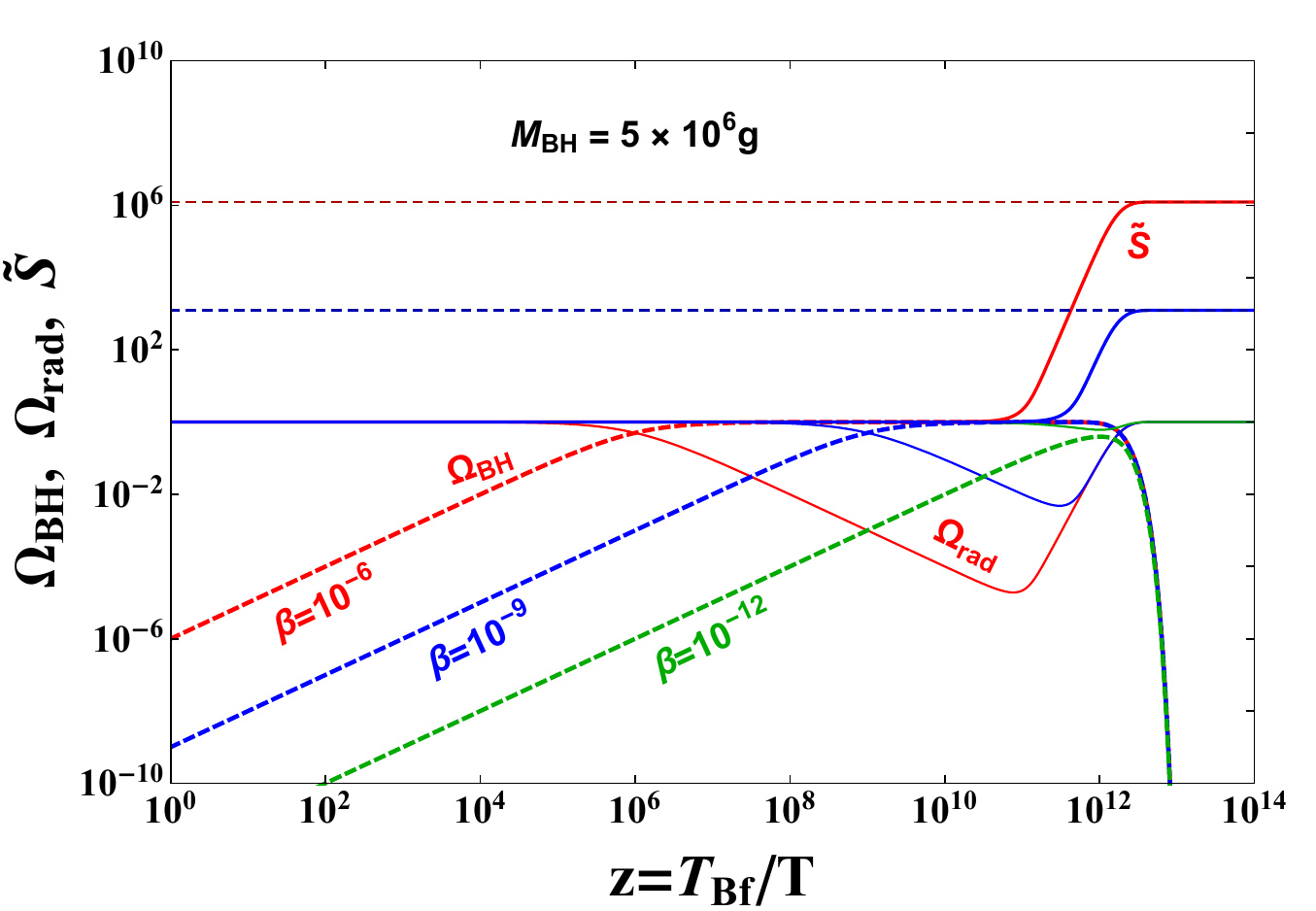}
\caption{Normalised energy densities as a function of inverse temperature (normalised to $T_{Bf}$). Because only the ratio of the total entropy $\tilde{S_i}\propto a_i^3/z_i^3$ are relevant, we have taken $\tilde{S}(z=1)=1$.}\label{fig2}
\end{figure}

The energy density of the black holes ($\rho_{\rm BH}$) and radiation ($\rho_R$) evolve according to the following Friedmann equations \cite{frd1,frd2}:
\bea
\frac{d\rho_{R}}{dz}+\frac{4}{z}\rho_R&=&0,\label{den1}\\
\frac{d\rho_{\rm BH}}{dz}+\frac{3}{z}\frac{H}{\tilde{H}}\rho_{\rm BH}&-&\frac{\dot{M}_{\rm BH}}{M_{\rm BH}}\frac{1}{z\tilde{H}}\rho_{\rm BH}=0,\label{den2}
\eea
where $H$ is the Hubble parameter and $z=T_{\rm Bf}/T$. The quantity $\tilde{H}$ and the scale factor $a$ evolve as
\bea
\tilde{H}=\left(H+\mathcal{K}\right),~
\frac{da}{dz}=\left(1-\frac{\mathcal{K}}{\tilde{H}}\right)\frac{a}{z},
\label{temvar}
\eea
where $\mathcal{K}=\frac{\dot{M}_{\rm BH}}{M_{\rm BH}}\frac{\rho_{\rm BH}}{4\rho_{R}}$. To derive Eq.\eqref{den1}-Eq.\eqref{temvar}, we assume the entropy ($g_{*s}$) as well as the energy ($g_{*\rho}$) degrees of freedom are equal and constant. For a given value of $\beta\equiv \frac{\rho_{\rm BH}(T_{\rm Bf})}{\rho_{\rm R}(T_{\rm Bf})}$,  the above equations can be solved to determine the duration of PBH domination and resulting entropy production $\Delta=\tilde{S}_{2}/\tilde{S}_{1}$, where $\tilde{S}_{1,2}\propto a_{1,2}^3/z_{1,2}^3$ is the total entropy before (after) the PBH evaporation.  In Fig.\ref{fig2}, we show the evolution of the normalised energy densities, $\Omega_{\rm BH,rad}=\rho_{\rm BH,rad}/\rho_{\rm BH}+\rho_{\rm rad}$, and consequent entropy production with $M_{\rm BH}=5\times 10^6$ g for three benchmark values of $\beta$.  From Fig.\ref{fig2}, it is evident that a larger value of $\beta$ corresponds to a longer period of PBH domination and larger entropy production. We find an analytical expression for $\Delta$ as\footnote{The entropy production factor can be well approximated as $\Delta\simeq \frac{3}{2}\frac{T_{\rm dom}}{T_{\rm eva}}$, where $T_{\rm dom}\simeq \beta T_{\rm Bf}$ is the temperature at which the PBHs start to dominate, and 3/2 is a numerical fitting factor that takes into account a finite duration of PBH evaporation.} 
\bea
\Delta \simeq 233~\beta\left(\frac{M_{\rm BH}}{M_{\rm Pl}}\right)\left(\frac{\gamma}{g_{*B}(T_{\rm BH})\mathcal{G}}\right)^{1/2},\label{delta}
\eea
where $\gamma\simeq 0.2$ is the PBH formation efficiency, $g_{*B}\simeq 100$ being the relativistic particles below $T_{\rm BH}$, and $\mathcal{G}\simeq 3.8$ is the greybody factor. Eq.\eqref{delta} matches the numerical solutions with very good accuracy, as shown in Fig.\ref{fig2} with the dashed horizontal lines. 

Like any other pre-existing relics, the produced baryon asymmetry from $N_1$ decays will be diluted due to the entropy production. Therefore, once the PBHs evaporate, we have the baryon-to-photon ratio:
\bea
\eta_B \simeq 10^{-2} N_{B-L}^{\rm final}=10^{-2} \Delta^{-1}\varepsilon_1 \kappa_1,\label{bas}
\eea
where $\varepsilon_1$ is the CP asymmetry parameter,  $\kappa_1$ is the efficiency of lepton asymmetry production, $N_{B-L}^{\rm final}$ is the final $B-L$ number, and the factor $10^{-2}$ accounts for the combined effects of sphaleron conversion of the leptons to baryons plus the photon dilution \cite{lep3}. We shall neglect the flavour effects \cite{flv1,flv2,flv3} in the leptogenesis computation,  use the maximum value of the CP asymmetry parameter, and work in the strong-washout regime of leptogenesis \cite{lep3,lep4}. Therefore, we consider the  CP asymmetry parameters and the efficiency factor as 
\bea
\varepsilon_1=\frac{3M_1  \sqrt{ |\Delta m_{\rm atm}|^2}}{8\pi v^2},~~\kappa_{1}\simeq 10^{-2},\label{lepasm}
\eea
where $v=174$ GeV is the vacuum expectation value of the SM Higgs, and $|\Delta m_{\rm atm }|^2\simeq 2.4\times 10^{-3}$ eV \cite{Esteban:2020cvm} is the active neutrino atmospheric mass-squared difference. Using Eq.\eqref{delta}, Eq.\eqref{lepasm}, and the observed value of $\eta_B\simeq 6.3\times 10^{-10}$ in Eq.\eqref{bas}, we obtain a simple analytic relation between $\beta$ and $M_1$ as 
\bea
\beta=5.7\times 10^{-12}\left(\frac{M_{\rm Pl}}{M_{\rm BH}}\right) M_1~{\rm GeV^{-1}},\label{betaM1}
\eea
which we shall use in the computation of GWs from PBHs.\\

There are several ways ultralight PBHs  involve in the production of GWs. For instance, the primordial curvature perturbations that produce PBHs also induce GWs (see, e.g., \cite{pbhgw-1,pbhgw0,pbhgw1}), PBHs radiate gravitons that constitute high-frequency GWs \cite{pbhgw2}, PBHs form mergers that emit GWs \cite{pbhgw3}, and finally, the inhomogeneous distribution of PBHs leading to density fluctuations, induces GWs \cite{ingw1,ingw2,ingw3}. We shall focus on the last one in this work. \\

It has been recently pointed out in Ref.~\cite{ingw1} and further developed in Refs.~\cite{ingw2,ingw3}, that right after formation, PBHs are randomly distributed in space according to Poisson statistics. Therefore, even though the PBH gas behaves as pressure-less dust on average, the spatially inhomogeneous distribution leads to density fluctuations, which are isocurvature in nature. Once the PBHs dominate the Universe's energy density, the isocurvature component gets converted to curvature perturbations that result in secondary GWs. Because the density fluctuations are large at small scales (comparable to the mean separation of PBHs at $T_{\rm Bf}$), sizeable GWs are induced, which are further enhanced due to the almost instantaneous (see Fig.\ref{fig2}) evaporation of PBHs \cite{ingw2,ingw4}. The present-day amplitude of such induced GWs is given by\footnote{Notice that the amplitude of the induced GWs is highly sensitive to the PBH mass spectrum \cite{ingw2,ingw4,Papanikolaou:2022chm}. Therefore, the results presented in this paper can vary significantly for an extended mass function instead of the monochromatic mass spectrum we use. Moreover, we consider only non-rotating Schwarzschild BHs. However, the evaporation temperature of a BH with spin (a Kerr BH) changes only slightly compared to a non-spinning BH \cite{kerr1,kerr2}, so that our results would remain essentially unchanged for spinning PBHs. In addition, let us mention that to derive Eq.(\ref{gul1}), we have used the most updated results presented in Refs.\cite{ingw2,ingw3,ingw4}. The initial version of Ref.\cite{ingw2} did not consider the suppression factor due to the finite width of the transition \cite{ingwm1}.}
\bea
\Omega_{\rm GW}(t_0,f)\simeq \Omega_{\rm GW}^{\rm peak}\left(\frac{f}{f^{\rm peak}}\right)^{11/3}\Theta
\left(f^{\rm peak}-f\right),\label{gul1}
\eea
where 
\bea
\Omega_{\rm GW}^{\rm peak}\simeq 2\times 10^{-6} \left(\frac{\beta}{10^{-8}}\right)^{16/3}\left(\frac{M_{\rm BH}}{10^7 \rm g}\right)^{34/9},\label{gul1s}
\eea
and
\bea
f^{\rm peak}\simeq 1.7\times 10^3 {\rm Hz}\left(\frac{M_{\rm BH}}{10^4 \rm g}\right)^{-5/6}.\label{gul2}
\eea
The $\Theta$-function in Eq.\eqref{gul1} stipulates that the Poisson spectrum of density fluctuation is subjected to an ultra-violet cut-off $f_{\rm UV}\simeq f^{\rm peak}$, which is comparable to the frequency corresponding to the comoving scale representing the mean separation of PBHs at the time of formation.
\begin{figure}
\includegraphics[scale=.5]{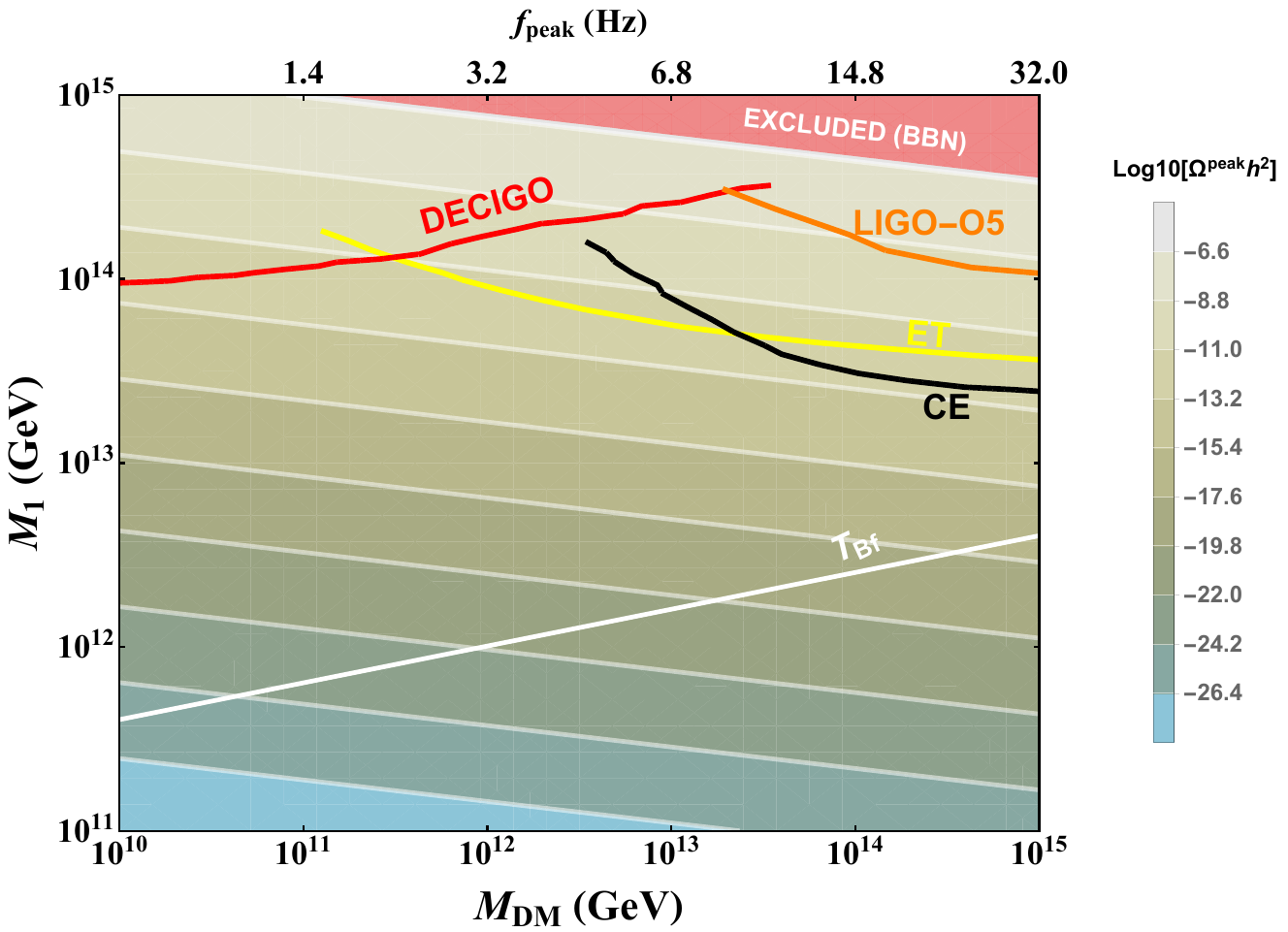}
\caption{Contours of $\Omega_{GW}^{\rm peak}$ on the $M_1-M_{DM}$ plane. The red-shaded region is excluded because otherwise the induced GWs would saturate the BBN bound on the effective number of neutrino species. The sensitivity limits of various experiments are shown with the coloured lines. The white line represents the PBH formation temperature, $T_{\rm Bf}\sim f\left(M_{\rm BH}\right)$, expressed as a function of DM mass (cf. Eq.(\ref{hdm})).}\label{fig3}
\end{figure}
Intriguingly, $\Omega_{\rm GW}^{\rm peak}$ being dependent on $\beta$ and $M_{\rm BH}$, can be expressed as a function of $M_1$ and $M_{\rm DM}$ (cf.\ Eq.\eqref{hdm} and Eq.\eqref{betaM1}) in this model. We derive the master equation relating the DM and scale of leptogenesis as
\bea
\Omega_{\rm GW}^{\rm peak}\simeq  \Omega_{\rm GW}^{\rm 0} \left(\frac{M_1}{10^{14}~\rm GeV}\right)^{16/3}\left(\frac{M_{\rm DM}}{10^{14}~\rm GeV}\right)^{28/45},\label{master1}
\eea
where $\Omega_{\rm GW}^{\rm 0}\simeq 2\times 10^{-10}$. Similarly,  $f_{\rm peak}$ can be expressed in terms of DM mass as
\bea
f^{\rm peak}\simeq 15~\left(\frac{M_{\rm DM}}{10^{14}~\rm GeV}\right)^{1/3} {\rm Hz}.\label{master2}
\eea
Although the amplitude depends on both $M_1$ and $M_{\rm DM}$, the peak frequency is determined only by $M_{\rm DM}$. Before we present the numerical results relevant to the GW experiments, let us point out a constraint that must be considered. Depending on $M_{\rm BH}$ and $\beta$, the induced GWs could be strong enough (Eq.\eqref{gul1s}) to saturate the BBN constraint on the effective number of neutrino species which is bounded from above \cite{neff}. Ref.~\cite{ingw2} derives an upper bound on $\beta$ as a function of $M_{\rm BH}$, which, using Eq.\eqref{betaM1} and Eq.\eqref{hdm}, we recast as 
\bea
M_{1,\rm max}^{\rm BBN}\simeq 4.6\times 10^{14}\left(\frac{M_{\rm DM}}{10^{14}~\rm GeV}\right)^{-7/60}~{\rm GeV}.\label{m1bbn}
\eea

In Fig.\ref{fig3}, we show the contours of $\Omega_{\rm GW}^{\rm peak}h^2$ on the $M_{\rm DM}-M_1$ plane. The BBN bound derived in Eq.(\ref{m1bbn}) excludes the red-shaded region. From this figure, we extract the following key points: 1) In a PBH-dominated early Universe, Fig.\ref{fig3} represents the most general testable parameter space in the seesaw with three super-heavy RH neutrinos, with one of them being the DM. The hierarchy in the masses could be in any order, i.e., $M_{\rm 3,DM}>M_{i=1,2}$ as well as  $M_{\rm 3, DM}<M_{i=1,2}$. For the former (latter) case, one expects to see a detectable GW signal at the higher (lower) frequencies. 2) Detectors such as CE, ET, and DECIGO can probe the model if the scale of leptogenesis is sufficiently high; $M_1\gtrsim 5\times 10^{13}$ GeV, and at the same time, if $f^{\rm peak}\gtrsim 0.6~ {\rm Hz}$. The absolute lower bound $f_{\rm min}^{\rm peak}\simeq 0.6~\rm Hz$ corresponds to the lowest value of the allowed DM mass ($M_{\rm DM}\simeq 10^{10}$ GeV) in the $\rm PBH\rightarrow DM $ mechanism. \\

The scenario becomes extremely predictive if the DM mass is known. In this model, the DM being super-heavy, we do not expect any effects on conventional DM searches \cite{Schumann:2019eaa}, even if we switch on DM interactions\footnote{An indirect way to measure super-heavy DM mass is to find the signatures in the cosmic rays \cite{cosmic1,cosmic2,cosmic3}.}. Recently,  Ref.~\cite{Samanta:2021mdm} pointed out that if the DM acquires mass by a  $U(1)$ gauge symmetry breaking, which also produces cosmic strings, a PBH-dominated Universe offers a unique way to determine the super-heavy DM mass from the spectral features of cosmic string-radiated GWs. An elegant UV completion of the seesaw consists in promoting the difference between the lepton ($L$) and the baryon ($B$) number to a new gauge symmetry $U(1)_{B-L}$, which can be naturally embedded in GUT \cite{bml1,bml2,bml3,bml4} -- one of our primary motivations as outlined in the introduction. In addition, a  three right-handed neutrino extension of the SM makes $G_{\rm SM} \times U(1)_{B-L}$ anomaly free ($G_{\rm SM}$ is the SM gauge group). In the next section, we discuss the $U(1)_{B-L}$ version of the seesaw in the presence of ultralight PBHs, namely the $U(1)_{B-L}^{\rm PBH}$ model.\\

\section{Gravitational waves signatures in the $U(1)_{B-L}^{\rm PBH}$ model}\label{sec3}
Cosmic strings \cite{cs1,cs2} appear as topological defects once the $U(1)_{B-L}$ breaks and the RH neutrinos become massive: $M_{i=1,2,3}= f_i v_\Phi$, where $f_i$ and $v_\Phi$ are the Yukawa coupling and  the VEV of the symmetry breaking scalar $\Phi_{B-L}$ respectively. After their formation, strings form closed loops and a network of horizon-size long strings \cite{ls1}. When two segments of the long-strings cross, they form loops. Long strings are characterized by a correlation length $L=\sqrt{\mu/\rho_\infty}$, where $\rho_\infty$ is the long-string energy density and $\mu $ is the string tension. For a gauge coupling $g$ and a scalar self-interaction coupling $\lambda$,  the string tension $\mu $ is defined as $\mu=\pi v_\Phi^2 \mathcal{H}\left( \frac{\lambda}{2 {g}^2}\right)$. The quantity $\mathcal{H}$ varies slowly \cite{book} with respect to its argument: $\mathcal{H}\left( \frac{\lambda}{2 {g}^2}\right)\simeq 1$ for $\lambda= 2  {g}^2$. As in the previous section, let us consider $N_3\equiv N_{\rm DM}$, i.e., $f_{i=3}\equiv f_{\rm DM}$, so that $ M_{\rm DM}=f_{\rm DM }v_\Phi$. Therefore, we can express the string tension in terms of DM mass as $\mu=\pi M_{\rm DM}^2 f_{\rm DM}^{-2}\mathcal{H}\left( \frac{\lambda}{2 {g}^2}\right)$.\\
 
Generally, owing to the strong interaction with the thermal plasma~\cite{Vilenkin:1991zk}, the motion of a string network gets damped. Once the damping becomes inefficient, the network oscillates to enter a scaling evolution phase, characterized by the fragmentation of the long strings into loops and stretching of the correlation length due to the cosmic expansion. These loops oscillate independently and produce GWs~\cite{Vachaspati:1984gt}.\\

Cosmic string network simulations find evidence of scaling solutions (see Refs.~\cite{scl1,scl2,scl3,scl4}), which motivates us to consider the network in the scaling regime in our computation. The size of a radiating loop  at a cosmic time $t$ is given by $l(t)=\alpha t_i-\Gamma G\mu(t-t_i)$, where $l_i=\alpha t_i$ is the initial size of the loop, $\Gamma\simeq 50$~\cite{Vilenkin:1981bx}, and $\alpha\simeq 0.1$ ~\cite{al1,al2}. The energy loss from a loop is decomposed into a set of normal-mode oscillations with frequencies $f_k=2k/l_k=a(t_0)/a(t)f$, where $k=1,2,3...\infty$.  The $k$th mode GW density parameter is obtained as \cite{al1} 
\bea
\Omega_{GW}^{(k)}(f)=\frac{2kG\mu^2 \Gamma_k}{f\rho_c}\int_{t_{osc}}^{t_0} \left[\frac{a(t)}{a(t_0)}\right]^5 n\left(t,l_k\right)dt,\label{gwf1}
\eea
where $n\left(t,l_k\right)$ is the  loop number density, and we  calculate  it using the Velocity-dependent-One-Scale (VOS) model \cite{vos1,vos2}.  The quantity $\Gamma_j$ is given by $\Gamma_j=\frac{\Gamma j^{-\delta}}{\zeta(\delta)}$, with $\delta=4/3$ for loops containing cusps \cite{Damour:2001bk}. Eq.\eqref{gwf1} is valid only for  $t_i > l_{\rm crit}/\alpha$, with $l_{\rm crit}$ the critical length above which GW emission dominates the massive particle radiation \cite{Matsunami:2019fss,Auclair:2019jip}, and $t_i>t_{\rm osc} = {\rm Max}\,\left[t_F,t_\text{fric}\right]$, where $t_F$ ($t_{\rm fric}$) is network formation (end of damping) time. \\

Detectable GWs from the cosmic string loops come from the most recent cosmic epochs. While the overall GW amplitude grows with $\mu$ \cite{al1}, in presence of a matter-era before the most recent radiation epoch at $T = T_\Delta$, the GW spectrum  at higher frequencies can be described as $\Omega_{\rm GW}^{(1)}(f\lesssim f_\Delta)\sim f^0 = \text{const}$  and  $\Omega_{\rm GW}^{(1)}(f\gtrsim f_\Delta)\sim f^{-1}$, with $f_\Delta$ the frequency of the spectral-break (see, e.g., Refs.\cite{br1,br2,br3,br4,br5,br6} for the importance of $f_\Delta$ to probe various cosmological models).\\

We now consider a simple scenario where $T_{\rm RH}\lesssim M_{\rm DM}$, so that the thermal bath does not have sufficient energy to produce the dark matter, despite the latter having gauge interaction, e.g.,  $g\simeq 1$. In this case, the $U(1)_{B-L}$ breaking must occur after the cosmic inflation ends. Otherwise, the inflated string network re-enters the horizon at late times, which may lead to other spectral breaks that can obfuscate the feature at $f_\Delta$ \cite{Guedes:2018afo}.\\

Using the PBH evaporation temperature in the case where they dominate \cite{frd1,frd2}  along with Eq.\eqref{hdm}, we obtain  $T_\Delta\equiv T_{\rm eva}^{\rm DM}=2.1\times 10^{-8} \left(M_{\rm DM}/{\rm GeV}\right)^{3/5}~{\rm GeV}$. Consequently,  an approximate analytical expression for $f_\Delta$\cite{br1,Samanta:2021mdm} in our case looks like
\bea
f_\Delta \simeq f_\Delta^0 \sqrt{\frac{50}{z_{\rm eq}\alpha \Gamma G\mu}}\left(\frac{\mu f_{\rm DM}^2}{\pi\mathcal{H} }\right)^{3/10}T_0^{-1}t_0^{-1},\label{dmbr}
\eea
where $f_\Delta^0\simeq 2.1\times 10^{-8}$, $t_0\simeq 4\times10^{17}$ sec,  $T_0=2.7 $K, $z_{\rm eq}\simeq 3387$. Because the dependence of Eq.\eqref{dmbr} on $\mathcal{H}$ is weak and $f_{\rm DM}\lesssim \sqrt{4\pi} $, remarkably, a GW measurement at low frequencies constrains $\mu$ while also robustly predicting an approximate upper bound on $f_\Delta$.\\

\begin{figure}
\includegraphics[scale=.5]{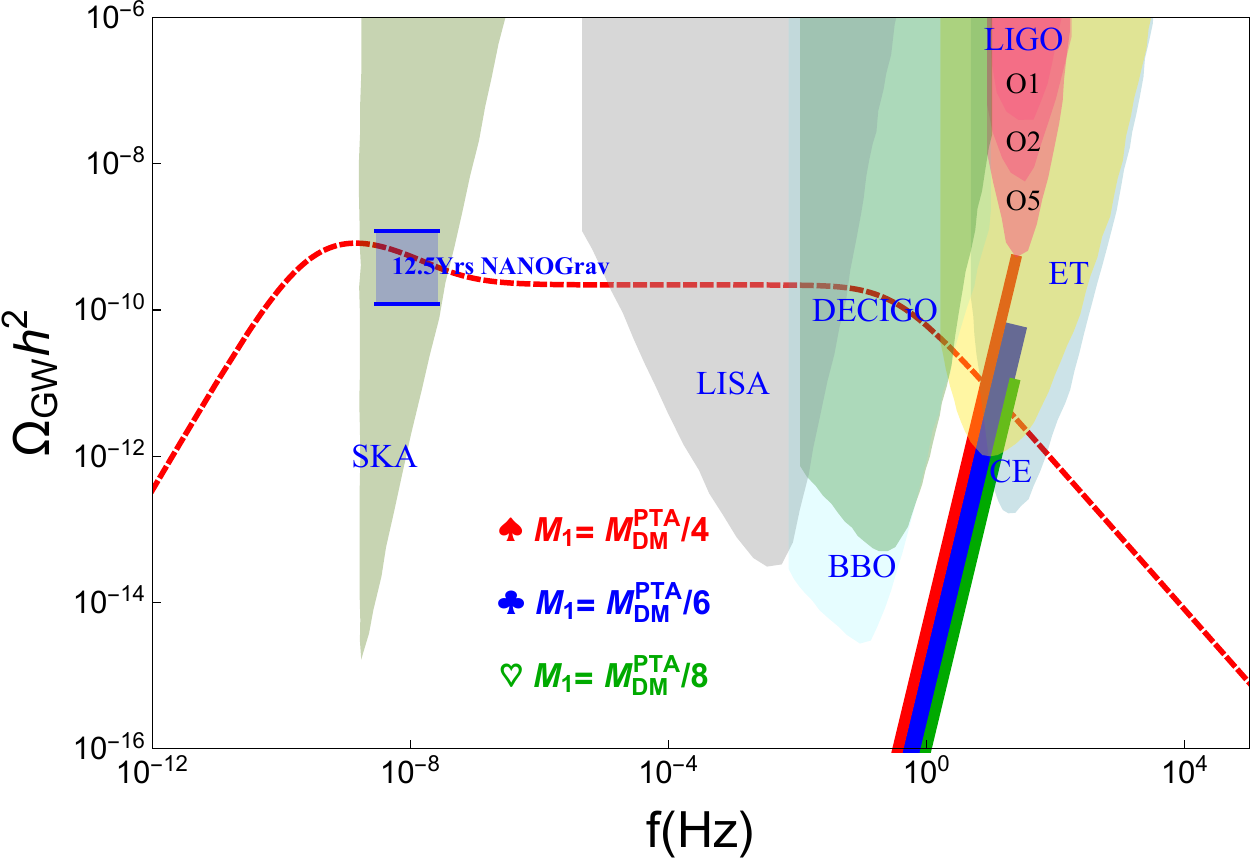}\\ 
\hspace{1.6cm}\includegraphics[scale=.5]{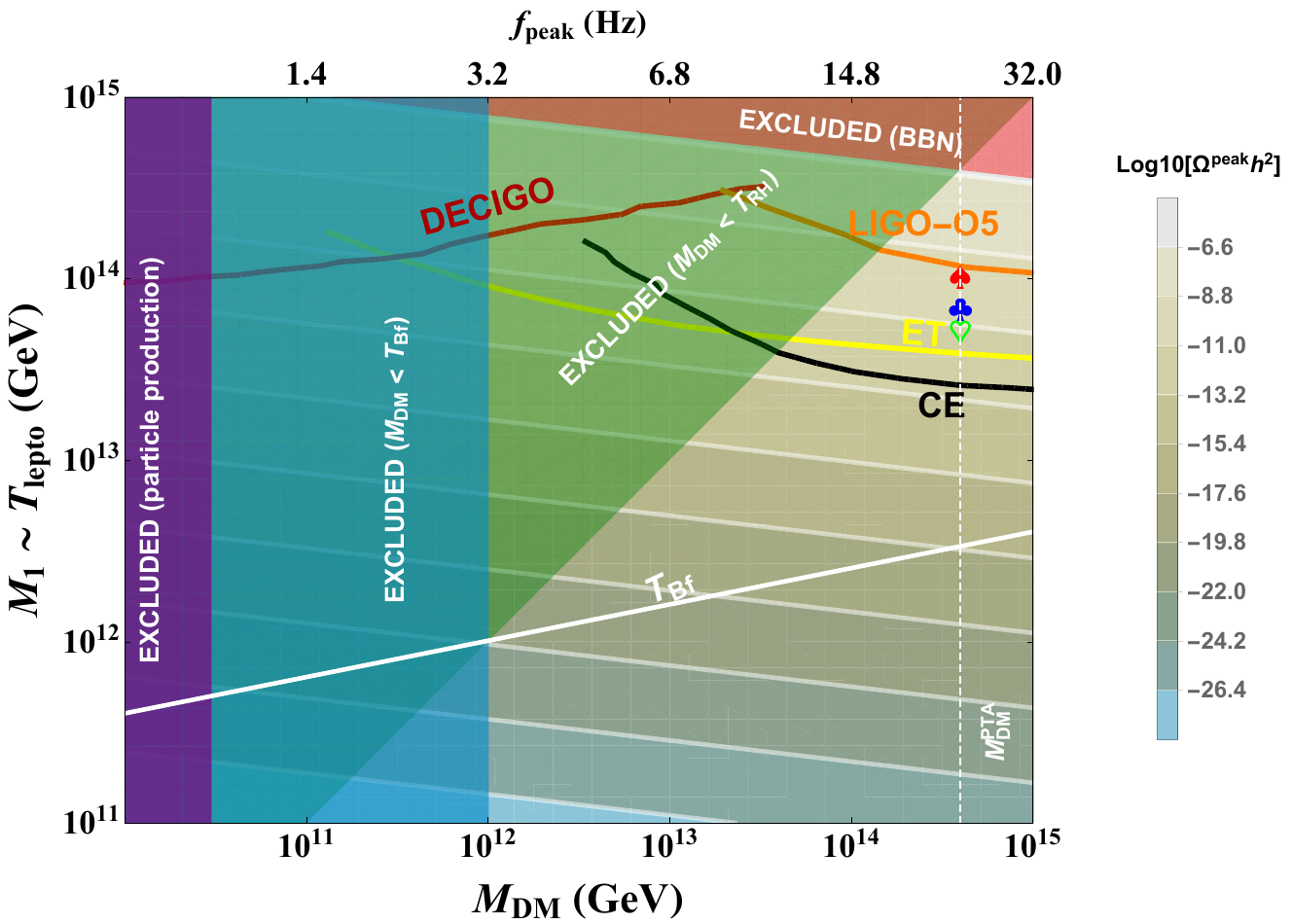}
\caption{Top: GW signatures of $U(1)_{B-L}$-seesaw in presence of ultralight PBHs. The dashed red line shows GWs from cosmic strings for $M_{DM}=4\times 10^{14}$ GeV. The slanted vertical bars show GWs from PBH density fluctuation corresponding to  $M_1=M_{DM}/x$, where $x=4~{\rm (red)},~6~{\rm (blue)}~ {\rm and}~ 8~{\rm (green)}$. The coloured regions represent the nominal sensitivities of several current and planned GW detectors as well as the NANOGrav result.  Bottom: contours of $\Omega_{GW}^{\rm peak}$ on the $M_1-M_{DM}$ plane, containing the benchmark cases shown on the left panel. See the text for the explanation of the exclusion regions.}\label{fig4}
\end{figure}

In this context, a significant experimental result at low frequencies is the recent finding of the NANOGrav pulsar-timing array experiment, which reports strong evidence of a stochastic common-spectrum process across 47-millisecond pulsars with 12.5 years of data \cite{ng}. However, the data set does not show the required evidence for quadrupolar spatial correlation described by Hellings-Downs curve \cite{hd}. Interestingly, such a common-spectrum process has been confirmed recently by the PPTA \cite{ppta}, and the EPTA collaboration \cite{epta}. The signal, if genuine, can be well explained in this model with $M_{\rm DM}^{\rm PTA}\simeq 3\times 10^{13}~{\rm GeV}- 6\times 10^{14}$ GeV assuming the DM maximally coupled to $\Phi_{B-L}$ and $\mathcal{H}\simeq 1$. Therefore, if the pulsar timing arrays results are due to the existence of a super-heavy DM with $M_{\rm DM}^{\rm PTA}\simeq \mathcal{O} (10^{14})$ GeV in the $U(1)^{\rm PBH}_{B-L}$ model, we forecast that DECIGO will observe a spectral break ($f_\Delta$) in GWs as shown in Fig.\ref{fig4} (top) with the dashed red curve. The signature is unique compared to the gravitational waves from cosmic strings always in radiation domination \cite{ncs1,ncs2,sp2} and compared to a cosmic string model in which there are other intermediate matter domination epochs, e.g., see Refs.~\cite{br1,br2,br3,br4,br5,br6}. For the former case, we do not expect any spectral break at higher frequencies, whereas, for the latter, $f_\Delta$ is generally not bounded.\\

Moving forward with a benchmark value $M_{\rm DM}^{\rm PTA}\simeq 4 \times 10^{14}$ GeV, we now  look for the signature of leptogenesis by substituting Eq.\eqref{master1} and Eq.\eqref{master2} in Eq.\eqref{gul1}. The results are shown with the coloured solid line for $M_1\simeq M_{\rm DM}^{\rm PTA}/x$, with $x=4$ (red), $x=6$ (blue), and $x=8$ (green) in Fig.\ref{fig4} (top)\footnote{In Fig.\ref{fig4} we show the GW spectrum radiated by cosmic strings for the $k=1$ mode. When we sum over an infinite number of modes, the spectrum behaves as $\Omega_{GW}(f>f_\Delta)\sim f^{-1/3}$ for loops containing cusps \cite{frd2}. Therefore, $M_{\rm DM}^{\rm PTA}/M_1$ should be bounded from above in order for induced GWs to dominate over GWs from cosmic strings beyond $f_\Delta$. We find that for $k\rightarrow \infty$, this translates to $M_{\rm DM}^{\rm PTA}/M_1\lesssim 6.5$. }. Note that the amplitude of the induced GWs increases with $M_1$. The reason being, a large $M_1$ produces more baryon asymmetry, which requires strong entropy dilution and hence large $\beta$--to be consistent with the observed value. Consequently, the induced GWs are enhanced according to Eq.\eqref{gul1s}. If $M_{i=1,2}$ are quasi-degenerate, the amount of asymmetry is greatly enhanced by the resonant leptogenesis mechanism \cite{Pilaftsis:2003gt}. In which case, even if $M_1\ll \Lambda_{\rm GUT}$, one can have induced GWs at the level of LIGO-5. Thus, in this mechanism, any measurement of GWs at low frequencies not only forecasts a spectral break in the string-radiated GWs but also robustly fixes the location of the peak of the induced GWs whose amplitude is determined by the scale of leptogenesis. Therefore, if timing array results are correct, in this model, we expect a signature of high-scale leptogenesis at mid-frequency interferometer scales, with $f^{\rm peak}\simeq 23.8 $ Hz.\\

In Fig.\ref{fig4} (bottom), we have reproduced the contours  of $\Omega_{\rm GW}^{\rm peak}$ as in Fig.\ref{fig3}, but now with three more exclusion regions and with the benchmarks shown in the left panel. The green region is excluded because we consider $M_{\rm DM}>T_{\rm RH}>M_{1,2}$. The blue region is excluded because otherwise, $T_{\rm Bf}>T_{\rm RH}$. In the less-stringent purple region, particle production is dominant over the gravitational radiation, where we have assumed that the string-loops contain only cusp-like structures \cite{Matsunami:2019fss} and $\mu\sim M_{\rm DM}^2$.\\

\begin{figure*}
\includegraphics[scale=.5]{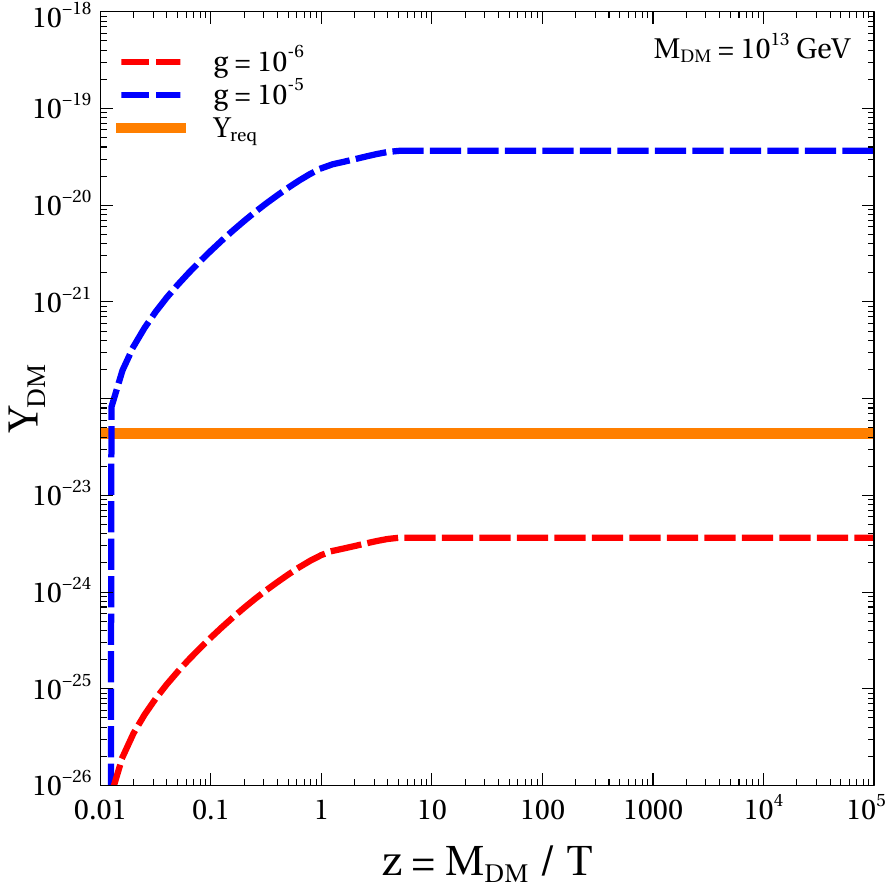} ~~~~ \includegraphics[scale=.5]{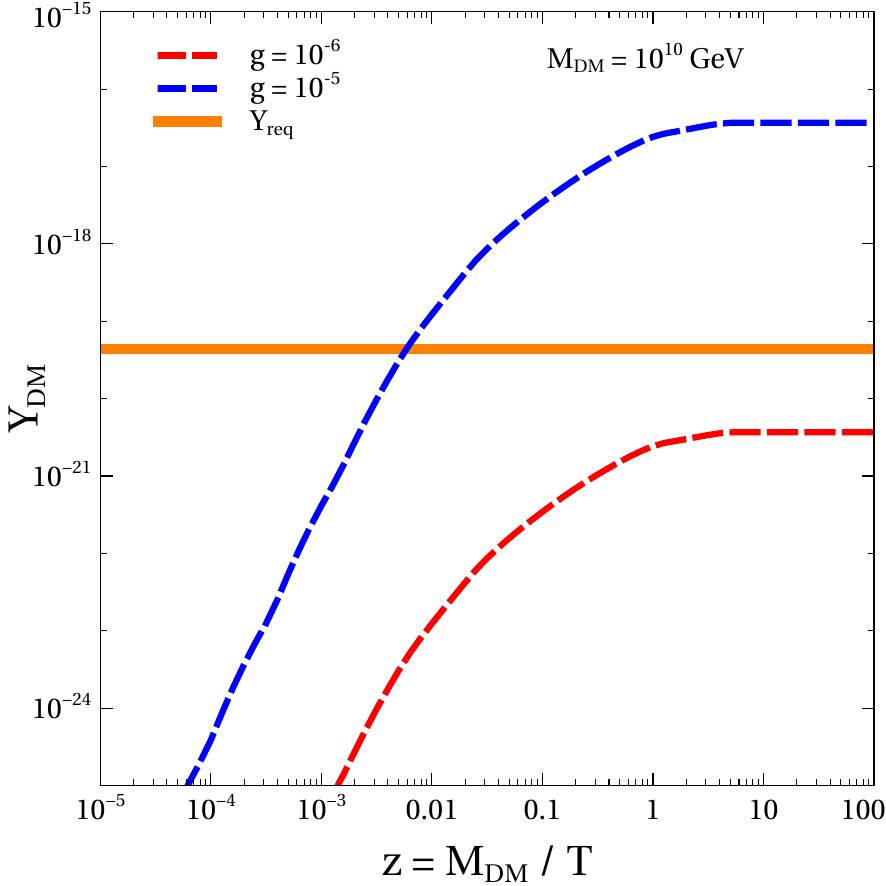}\\
\caption{Evolution of DM yield (normalised to entropy density) produced from freeze-in for $M_{\rm DM}=10^{13}$ GeV (left panel) and $M_{\rm DM}=10^{10}$ GeV (right panel). The red (blue) dashed lines are for $g=10^{-5}$ ($g=10^{-6}$). The orange contour denotes the yield required to get the observed DM abundance. We have taken $T_{\rm RH}=10^{15}$ GeV and the symmetry breaking scale $v_\Phi=10^{13}$ GeV.}\label{fig5}
\end{figure*}

A similar analysis for the $T_{\rm RH}>M_{\rm DM}$ can be done straightforwardly. Nonetheless, in this case, one needs to identify the domain of $g$ for which dark matter production via gauge boson-mediated interactions is suppressed. Otherwise, Eq.\eqref{dmbr}, which accounts only for $\rm PBH\rightarrow DM$ channel, will not be valid. For large values of $g$, which numerical simulations of cosmic string network take into account \cite{Matsunami:2019fss}, super-heavy dark matter will overclose the Universe via a thermal freeze-out mechanism \cite{Griest:1989wd}. Therefore, a suitable domain of $g$, in this case, is the freeze-in regime with small values of $g$ \cite{Hall:2009bx}. In this regime, though the dark matter never thermalizes, it can be produced via $B-L$ gauge boson mediated scatterings: $f \overline{f} \leftrightarrow N_{\rm DM}N_{\rm DM}$, where $f$ is an SM fermion. In Fig. \ref{fig5}, we show the evolution of the DM yields for two benchmark masses. We find that for $g \lesssim 10^{-6}$, the production channels in the freeze-in regime remain subdominant compared to $\rm PBH\rightarrow DM$ for the entire range of dark matter mass we are interested in. Therefore, for $T_{\rm RH}>M_{\rm DM}$, a predictive cosmic string phenomenology requires $g\lesssim 10^{-6}$.\\

Let us outline the following important characteristics of this case. I) The green and blue shaded exclusion regions are less important because $T_{\rm RH}$ is the highest scale in the theory. II) One can explore three limiting cases: $\lambda=2 g^2$, $\lambda\gg 2 g^2$ and $\lambda \ll 2 g^2$. Taking into account the results of numerical simulations with large coupling constants at face value, the first choice would provide a similar phenomenology as in $M_{\rm DM}>T_{\rm RH}$ case because $\mathcal{H}\simeq 1$. On the other hand, for the other two cases, one has $\mathcal{H}\simeq {\rm ln}\left(\frac{\lambda}{2g^2}\right)$ and $\mathcal{H}\simeq \left({\rm ln}~\frac{2g^2}{\lambda}\right)^{-1}$, respectively \cite{book}. Despite $\mathcal{H}$ having a logarithmic dependence on the parameters, these two cases may be less appealing because in Eq.\ref{dmbr} $\mathcal{H}$ appears as a free parameter. III) For small values of $\lambda$, the string-width $\delta_{\omega}\sim 1/\sqrt{\lambda} v_{\Phi}$ might constitute a considerable fraction of the horizon $H(T_F)^{-1}$ at initial times, assuming $v_\Phi$ is of the order of the network formation temperature, $T_F$. In this case, it is questionable to treat the strings as Nambu-Goto-strings ($\delta_{\omega}\ll H(T)^{-1} $) consistently throughout the evolution of the Universe. Notice that we consider the effect of PBHs on the cosmic string network only at the level of the Universe's expansion. We have ignored the possible interactions of PBHs and strings \cite{Vilenkin:2018zol} that might cause further spectral distortions. This requires numerical simulations of a PBH-string network, which is beyond the scope of this work.
\section{conclusion}
We have shown that the presence of ultralight black holes ($M_{\rm BH}<10^9$ g) in the early Universe may open up new avenues to study the cosmological implications of seesaw models. In most unified scenarios of dark matter and baryogenesis via leptogenesis the right-handed neutrinos are light, so that the theory is predictive for particle physics experiments. In this work, we showed that super-heavy right-handed neutrinos can be just as predictive. However, in this case, the search strategy for signatures is different. The presence of black holes and a super-high-scale phase transition that generates right-handed neutrino masses offer unique gravitational wave signatures that are testable in the upcoming gravitational wave detectors. In our model, ultralight black holes generate super-heavy right-handed neutrino dark matter via Hawking radiation, whereas baryogenesis is realized via thermal leptogenesis through the remaining pair of right-handed neutrinos. The amplitude of gravitational waves from PBH density fluctuations correlates with super-heavy neutrino masses. The detectability of such gravitational waves improves as the right-handed neutrino masses approach the grand unification scale; $\Lambda_{\rm GUT}$. In a realistic version of the model, motivated by grand unified theories, the presence of cosmic strings make the gravitational wave signatures distinct from other realisations of this idea, therefore offering a unique avenue to test its properties. 

\section*{acknowledgement}
We thank Guillem Domènech for very useful comments and suggestions. The work of RS is supported  by the  project International mobility MSCA-IF IV FZU - CZ.02.2.69/0.0/0.0/$20\_079$/0017754 and Czech Science Foundation, GACR, grant number 
20-16531Y. RS also acknowledges the European Structural and Investment Fund and the Czech Ministry of Education, Youth, and Sports. FU is supported by the European Regional Development Fund (ESIF/ERDF) and the Czech Ministry of Education, Youth and Sports (MEYS) through Project CoGraDS - CZ.02.1.01/0.0/0.0/$15\_003$/0000437.

\appendix
\section{Derivation of the particle production cut-off (lower bound on the DM mass)}
Gravitational waves production is dominant when $t_{\rm crit} >\frac{l_{\rm crit}}{\alpha}$, where for strings containing cusps-like structures we have $l_{\rm crit}=\frac{\mu^{-1/2}}{(\Gamma G \mu)^2}$  \cite{Matsunami:2019fss,Auclair:2019jip}. For the spectral break to happen due to PBH domination, the condition $t_{\rm eva}>t_{\rm crit}$ must be satisfied. This translate to a lower bound on $G\mu$:
\bea
G\mu > T_{\rm eva}^{4/5}\left(2.9\times 10^{-20}\right)^{4/5}.\label{apeq1}
\eea
Given that $T_{\rm eva}\simeq 2.8\times 10^{-8}\left(\frac{M_{\rm DM}}{\rm GeV}\right)^{3/5}$, Eq.(\ref{apeq1}) translates to 
\bea
G\mu> 2.1\times 10^{-22}\left(\frac{M_{\rm DM}}{\rm GeV}\right)^{12/25}.
\eea
Assuming $\mu\simeq \pi M_{\rm DM}^2$, we obtain a lower bound (which we show as the $M_{\rm DM}<M^{\rm min}_{\rm DM}$ region with purple color in Fig.\ref{fig4}) on the dark matter mass as
\bea
M_{\rm DM}\gtrsim 3\times 10^{10}~{\rm GeV}.
\eea
{}
\end{document}